\documentclass[%
reprint,
superscriptaddress,
amsmath,amssymb,
prc,
floatfix,]%
{revtex4-1}

\usepackage{CJK}
\usepackage{graphicx}
\usepackage{dcolumn}
\usepackage{bm}
\usepackage{enumerate}
\usepackage[bookmarks=true,colorlinks=true,linkcolor=blue,
           citecolor=blue,hypertex,breaklinks=true,urlcolor=blue]{hyperref}
\usepackage{booktabs}
\usepackage{array}
\usepackage{multirow}
\usepackage{xcolor,amsmath}
\usepackage{longtable}

\begin{document}

\title{Comparison of several model averaging methods in nuclear charge radius predictions}

\begin{CJK*}{GBK}{}

 \author{Huan-Yu Zhang}
 \affiliation{Mathematics and Physics Department,
              North China Electric Power University, Beijing 102206, China}

 \author{Rui Jing}
 \affiliation{Mathematics and Physics Department,
              North China Electric Power University, Beijing 102206, China}

 \author{Zhen-Hua Zhang}
 \email{zhzhang@ncepu.edu.cn}
 \affiliation{Mathematics and Physics Department,
              North China Electric Power University, Beijing 102206, China}
 \affiliation{Hebei Key Laboratory of Physics and Energy Technology,
              North China Electric Power University, Baoding 071000, China}

 \author{Xin-Hui Wu}
 \affiliation{Department of Physics, Fuzhou University, Fuzhou 350108, China}

 \author{Zhong-Ming Niu}
 \affiliation{School of Physics, Anhui University, Hefei 230601, China}

\date{\today}

\begin{abstract}
The performance of five model averaging methods, including the arithmetic mean (AM), weighted mean (WM),
naive Bayesian model averaging (NBMA), principal component analysis (PCA), and power-moderated mean (PMM) methods,
in nuclear charge radius predictions is investigated.
Five commonly used nuclear charge radius models are adopted as inputs for the averaging procedures.
The charge radius differences between the experimental data and the original nuclear models are
analyzed and the results after considering the model averaging methods are also discussed.
The calculations show that the NBMA method can provide the best root-mean-square (rms) deviation among these five model averaging methods.
The PCA method can extract useful physical information and
not only helps to interpret the model differences but also offers
a feasible way to construct improved empirical models by recombining the principal components.
In contrast to the other methods, whose results worsen upon including a new model with a larger
rms deviation, the rms deviation of the PCA method remains almost unaffected.
The PMM method is capable of integrating the strengths of various nuclear models
and delivering reasonable uncertainty estimates not only in known regions but also in unknown ones.
This method can automatically adjust data uncertainties to achieve consistency,
and it can provide a tool for a smooth transition of the nuclear charge radius prediction from the WM to the AM.
The extrapolation ability of these model averaging methods is checked by 66 newly observed data after year 2021.
The calculations show that model averaging offers a reliable strategy for nuclear charge radius predictions,
combining high accuracy on known data with robust extrapolation to new measurements.
The choice of the optimal averaging method may depend on the specific application scenario:
PCA and NBMA are preferred when the test set is expected to resemble the training distribution,
whereas PMM is more suitable for predictions in unexplored regions.
The charge radii and the odd-even staggering in calcium isotopes are also discussed.
The results indicate that if nearly all underlying models fail to reproduce the fine structure of nuclear charge radii,
the model averaging methods cannot correct this deficiency either.
\end{abstract}

\maketitle
\end{CJK*}

\section{Introduction}\label{sec:intro}

The charge radius ranks among the most fundamental properties of the atomic nucleus,
and its knowledge is essential for interpreting a wide range of nuclear phenomena,
including shell structure~\cite{Angeli2015_JPG42-055108, Gorges2019_PRL122-192502},
shape transitions~\cite{Wood1992_PR215-101, Cejnar2010_RMP82-2155},
neutron skin and halo~\cite{Tanihata1985_PRL55-2676, Tanihata2013_PPNP68-215, Meng2015_JPG42-093101}, among others.
Therefore, reliable predictions of nuclear charge radii are of great significance.
Thanks to continuous improvements in experimental techniques,
substantial progress has been made in the measurement of nuclear charge radii.
Over 1000 experimental values~\cite{Angeli2013_ADNDT99-69, Li2021_ADNDT140-101440}
have now been determined via diverse methods, such as electron elastic scattering,
muonic atom $X$-ray spectroscopy, and isotope shift measurements~\cite{Cheal2010_JPG37-113101, Campbell2016_PPNP86-127}.
Nevertheless, this number remains inadequate when contrasted with the vast number of nuclei
predicted to exist by various theoretical models~\cite{Goriely2001_ADNDT77-311, Erler2012_Nature486-509,
Moller2016_ADNDT109-110-1, Xia2018_ADNDT121-122-1, Guo2024_ADNDT158-101661}.

Various theoretical methods have been developed to describe charge radii,
including phenomenological formulae~\cite{Bohr1969_Book, Zeng1957_APS13-357, Nerlo-Pomorska1993_ZPA344-359,
Duflo1994_NPA576-29, Zhang2002_EPJA13-285, Lei2009_CTP51-123, Wang2013_PRC88-011301,
Bayram2013_APPB44-1791, Sheng2015_EPJA51-40, Jiao2025_MPLA40-2450216},
local-relation-based models~\cite{Piekarewicz2010_EPJA46-379, Sun2014_PRC90-054318,
Bao2016_PRC94-064315, Sun2017_PRC95-014307, Bao2020_PRC102-014306,
Ma2021_PRC104-014303, Li2023_ChinPhysC47-084104, Bao2023_NPR40-141},
macroscopic-microscopic models~\cite{Buchinger1994_PRC49-1402, Buchinger2001_PRC64-067303,
Buchinger2005_PRC72-057305, Iimura2007_PRC76-057302, Iimura2008_PRC78-067301}, as well as
non-relativistic~\cite{Fayans2000_NPA676-49, Stoitsov2003_PRC68-054312, Goriely2009_PRL102-242501,
Goriely2010_PRC82-035804, Reinhard2017_PRC95-064328, Reinhard2021_PRC103-054310,
Naito2023_PRC107-054307, Inakura2024_PRC110-054315}
and relativistic~\cite{Lalazissis1999_ADNDT71-1, Geng2005_PTP113-785, Zhao2010_PRC82-054319,
Xia2018_ADNDT121-122-1, Zhang2020_PRC102-024314, An2020_PRC102-024307,
Perera2021_PRC104-064313, Zhang2022_ADNDT144-101488, Zhang2023_PRC108-024310, Naito2023_PRC107-054307,
Xie2024_PRC110-064319, Xie2025_PRC112-L021303, Pan2025_PRC112-024316} mean-field models.
These approaches have been widely adopted for systematic investigations of nuclear charge radii.
In addition, the $ab$ initio calculations have also been applied to this topic~\cite{Forssen2009_PRC79-021303,
Choudhary2020_PRC102-044309, Elhatisari2024_Nature630-59, Ren2025_PRL135-152502, Sun2025_PRC112-024317}.
Generally, these models can provide overall satisfactory descriptions across the nuclear chart.
However, except for those based on local relations, the root-mean-square (rms) deviations tend to be somewhat larger,
and the description of fine structures, such as the evolution of charge radii
in calcium isotopes~\cite{Ruiz2016_NatPhys12-594, Miller2019_NatPhys15-432}
and the odd-even staggering in $N=94$ isotones~\cite{Takacs2025_arxiv2511.19395, Staiger_arXiv2511.20537},
still remains challenging.

In recent years, machine learning methods have been extensively employed across various domains of
nuclear physics~\cite{Bedaque2021_EPJA57-100, Gao2021_NST32-109, Boehnlein2022_RMP94-031003,
He2023_NST34-88, He2023_SciChinaPAM66-282001}.
To further improve the description of nuclear charge radii, a variety of machine learning approaches have been applied,
including artificial neural networks~\cite{Akkoyun2013_JPG40-055106, Wu2020_PRC102-054323,
Bayram2023_PS98-125310, Yang2023_PRC108-034315},
Bayesian neural networks~\cite{Utama2016_JPG43-114002, Neufcourt2018_PRC98-034318, Ma2020_PRC101-014304,
Dong2022_PRC105-014308, Dong2023_PLB838-137726, Li2025_PRC112-014312, Xian2025_PLB868-139662, Yuan2026_EPJA62-111},
the radial basis function method~\cite{Li2023_NPR_40-31, Li2026_ChinPhyC50-054102, Li2026_PRC114-014333},
the support vector regression~\cite{Jalili2024_NJP26-103017},
convolutional neural networks~\cite{Su2023_Symmetry15-1040, Cao2023_NST34-152},
CatBoost and XGBoost regression~\cite{Ahmed2026_ChinPhysC50-024109, Meng2026_PRC114-024314},
and kernel ridge regression~\cite{Ma2022_CPC46-074105, Tang2024_NST35-19, Tang2024_NSR41-927}, etc.
By training on the radius residuals (defined as the differences between experimental and calculated charge radii),
these methods can reduce the corresponding rms deviations to around 0.01-0.02~fm.
In addition, several works have focused on predicting nuclear charge density distributions via
neural networks~\cite{Shang2022_NST33-153, Shang2024_PRC110-014308, Guo2025_PS100-125308, Wang2026_NST37-93},
from which the charge radii can then be extracted.

Although machine learning methods have achieved high accuracy in predicting nuclear charge radii,
their extrapolation capabilities warrant careful verification,
as these models typically involve a large number of parameters.
Their predictions of nuclear charge radii also generally suffer from large uncertainties,
particularly in regions far from the known training data.
Using the kernel ridge regression method as a case in point, the extrapolation step toward the neutron-rich region is generally
less than 3 for the majority of nuclear models~\cite{Ma2022_CPC46-074105, Tang2024_NST35-19, Tang2024_NSR41-927}.
Model averaging approaches can improve robustness by combining predictions from multiple theoretical
nuclear charge radius models, thereby incorporating the underlying physics embedded in each model.
Among these, the arithmetic mean (AM) and weighted mean (WM) are the simplest and most frequently employed methods.
However, the AM method assigns identical weights to all models,
thus completely neglecting the uncertainties of individual models,
a practice that is typically justified only in extreme cases where the uncertainties are entirely unreliable.
In contrast, the WM method uses the reciprocals of the squared standard uncertainties as weights,
which is appropriate in the ideal scenario of a consistent dataset with well-determined uncertainties.
In practice, however, the reliability of these uncertainties is not always guaranteed, and they are often underestimated.

In recent years, numerous model averaging methods have been proposed and are now widely used in nuclear physics.
Different methods often yield different results, mainly owing to their varying definitions of ``optimality'',
criteria for weight assignment, and approaches to uncertainty treatment.
Hence, there is no universally optimal method; its actual performance depends heavily on
the data characteristics and the candidate model set,
and should therefore be chosen flexibly according to the specific problem.

The Bayesian model averaging (BMA) method constitutes a classical approach
for addressing model uncertainty within the Bayesian paradigm~\cite{Hoeting1999_SS14-382}.
Its fundamental principle involves weighting the predictions of all candidate models
by their respective posterior probabilities, thereby synthesizing information across multiple models.
The weights are jointly determined by the prior model probabilities and the marginal likelihoods,
reflecting the relative plausibility of each model conditional on the observed data.
BMA not only enhances predictive robustness but also naturally quantifies the uncertainty inherent in model selection.
It has been widely used in various branches of nuclear physics~\cite{Neufcourt2019_PRL122-062502,
Neufcourt2020_PRC101-044307, Neufcourt2020_PRC101-014319, Saito2024_PRC109-054301,
Alhassan2024_NST35-205, Xie2024_PRC109-064317, Liu2025_NST36-215}.
Nevertheless, its successful implementation depends crucially on reasonable prior specifications,
a comprehensive set of candidate models, and accurate computation of posterior probabilities.
By assuming that the events are independent of each other, the naive Bayesian probability formula can be used
to simplify the Bayesian model averaging method, which is called the naive Bayesian model averaging (NBMA) method.
This method has been applied to nuclear mass predictions in Ref.~\cite{Zhang2024_NPA1043-122820} and has achieved great success.

Principal component analysis (PCA) is a classical unsupervised linear dimensionality reduction technique
that transforms a set of potentially correlated observables into a few uncorrelated principal components (PCs),
capturing the dominant variance structure of the data~\cite{Wold1987_CILS2-37, Jolliffe2002_Book}.
In nuclear physics, PCA has proven particularly useful for handling high-dimensional datasets,
which are common when numerous theoretical models or experimental measurements produce highly correlated
outputs across a wide range of nuclear observables~\cite{Bulgac2018_PRC97-044313,
Fox2020_PRC101-054308, Kejzlar2020_JPG47-094001, Schunck2020_JPG47-074001,
Akimov2022_PRL129-081801, Bonilla2022_PRC106-054322, Wu2022_PRC105-L031303,
Augier2023_PRL131-162501, Wu2024_SSPMA67-272011, Lv2026_PLB874-140283}.
By reducing redundancy and distilling the key patterns, PCA facilitates model intercomparison,
outlier detection, and feature extraction, while also aiding
in the visualization of global trends across the nuclear chart.
Moreover, its ability to identify hidden correlations among different observables makes
it a valuable tool for data-driven analyses and for constructing empirical indicators that
may inform nuclear structure and reaction studies.
However, the linear nature of PCA and its sensitivity to data scaling should be borne in mind;
proper standardization of input variables is therefore generally recommended before application,
especially when combining data of different physical dimensions.

The power-moderated mean (PMM) method is a novel model averaging technique proposed to derive a key comparison
reference value and its associated standard uncertainty~\cite{Pomme2012, Pomme2015_Metrologia52-S200}.
It offers several advantages: automatic adjustment of data uncertainties to ensure consistency,
statistically based correction or exclusion of extreme data,
and reduction of the influence of uncertainties in weighting factors-yielding
more realistic uncertainty estimates for small datasets.
As a result, the PMM method provides efficient and robust means for both mutually consistent
data and datasets suspected of inconsistency~\cite{Pomme2012, Pomme2015_Metrologia52-S200}.
In the context of nuclear charge radius models, uncertainties are typically estimated from the
rms deviations between theoretical predictions and experimental data.
While these estimates are reasonable in regions where experimental data are available,
they are likely underestimated in unexplored regions.
The PMM method is capable of integrating the strengths of various nuclear models
and delivering reasonable uncertainty estimates not only in known regions but also in unknown ones.
This method has already been successfully applied to nuclear mass predictions
in Ref.~\cite{Zhang2024_PRC110-044307}, achieving notable success.

The objective of this study is to assess the advantages and disadvantages of several
model averaging techniques in the context of nuclear charge radius descriptions.
We benchmark their predictive performance using five selected methods,
namely AM, WM, NBMA, PCA, and PMM.
Section~\ref{sec:theor} briefly introduces the frameworks of the NBMA, PCA, and PMM methods,
while the numerical details of the calculations are described in Sec.~\ref{sec:details}.
The results from these averaging approaches are presented in Sec.~\ref{sec:result},
and a brief summary is given in Sec.~\ref{sec:summary}.

\section{Theoretical framework}\label{sec:theor}

This section concisely reviews prevalent model averaging methodologies in nuclear structure studies:
(i) naive Bayesian model averaging~\cite{Zhang2024_NPA1043-122820},
(ii) principal component analysis~\cite{Wu2024_SSPMA67-272011}, and
(iii) power moderated mean~\cite{Zhang2024_PRC110-044307}.
For comparative benchmarking, the arithmetic mean method and weighted mean method are implemented,
with their respective nuclear charge radius predictions given by
\begin{align}
\overline{R}_{\mathrm{AM}} &= \frac{1}{K} \sum_{i=1}^{K} R_{Z,N}^i, \label{eq:amm} \\
\overline{R}_{\mathrm{WM}} &= \sum_{i=1}^{K} \omega_i R_{Z,N}^i, \label{eq:wam}
\end{align}
where $K$ denotes the number of theoretical models,
and $\omega_i$ represents the normalized probability weight for model $i$,
quantified via the root-mean-square (rms) deviation between model predictions and experimental charge radii.

\subsection{Naive Bayesian model averaging method}

Within the BMA framework, the posterior distribution $P(R_{Z,N}|D_{Z,N})$ for the
nuclear charge radius $R_{Z,N}$ conditioned on the experimental dataset $D_{Z,N}$ can be written as
\begin{eqnarray}
P(R_{Z,N}|D_{Z,N}) = \sum_{k=1}^{K} P(R_{Z,N}|M^k) P(M^k|D_{Z,N}),
\end{eqnarray}
where $M^1$, $M^2$, $\cdots$, $M^K$ denote the individual models in the ensemble under consideration.
The predictive probability density function $P(R_{Z,N}|M^k)$,
derived solely from model $M^k$, takes the form of a Dirac delta function $\delta(R - R_{Z,N}^k)$.
Here $R_{Z,N}^k$ represents the charge radius prediction for nucleus $(Z,N)$ by model $M^k$.
The posterior model probability $P(M^k|D_{Z,N})$, quantifying $M^k$'s agreement with experimental data $D_{Z,N}$,
is calculated by Bayes' theorem as
\begin{eqnarray}
P(M_k|D_{Z,N}) = \frac{P(D_{Z,N}|M^k) P(M^k)}{\sum_{j=1}^{K} P(D_{Z,N}|M^j) P(M^j)}.
\end{eqnarray}
where $P(D_{Z,N}|M^k)$ denotes the likelihood function and $P(M^k)$ represents the prior probability for model $M^k$.
The BMA prediction is then given by the posterior mean of the nuclear charge radius, formally expressed as
\begin{eqnarray}
\overline{R}_{Z,N} = \int R_{Z,N} P(M_k|D_{Z,N}) dR_{Z,N}.
\end{eqnarray}

In the NBMA method, the likelihood can be calculated by supposing
\begin{eqnarray}
P(D_{Z,N}|M^k) \propto P(D_{Z}|M^k) P(D_{N}|M^k),
\end{eqnarray}
where the likelihood functions $P(D_Z|M^k)$ and $P(D_N|M^k)$
are quantified using the rms deviation between model $M^k$'s charge radius predictions
and experimental data for nuclei with proton number $Z$ and neutron number $N$, respectively.
The normalized posterior probabilities are denoted as $P(Z|M^k)$ and $P(N|M^k)$.
Given the absence of compelling evidence favoring any specific model,
an uninformative prior $P(M^k)$ is adopted for all $K$ candidate models.
Consequently, the NBMA charge radius prediction is calculated by
\begin{eqnarray}
\overline{R}_{Z,N} = \sum_{k=1}^{K} R_{Z,N}^k P(Z|M^k) P(N|M^k).
\end{eqnarray}

\subsection{Principal component analysis method}

The application of PCA to nuclear charge radius models proceeds as follows:

\textbf{1. Model Selection:} first,
select $K$ charge radius models for analysis
\begin{equation}
\mathcal{R}_1, \mathcal{R}_2, \dots, \mathcal{R}_K.
\end{equation}

\textbf{2. Data Vectorization:} second,
transform each model's charge radius predictions into $m$-dimensional vectors
\begin{equation}
\mathbf{R}_k = [R_k^{1}, R_k^{2}, \dots, R_k^{m}]^{\top},
\end{equation}
where $m$ corresponds to the number of nuclei in the nuclear chart,
and these $m$ nuclei are the overlap of these charge radius models.
These constitute the original charge radius model vectors.

\textbf{3. Covariance Construction:} third,
form the $K \times m$ data matrix $\mathbf{X} = [\mathbf{R}_1, \mathbf{R}_2, \dots, \mathbf{R}_K]^{\top}$ and compute the $m \times m$ covariance matrix:
\begin{equation}
\mathbf{C} = \frac{1}{m} \mathbf{X}^{\top}\mathbf{X}.
\end{equation}
The rank of $\mathbf{C}$ is $K$.

\textbf{4. Eigen-Decomposition:} fourth,
diagonalize $\mathbf{C}$ to obtain eigenvalues $\lambda_i$ and eigenvectors
$\boldsymbol{\nu}_i$ (the \textit{principal components}),
ordered such that $\lambda_1 \geq \lambda_2 \geq \cdots \geq \lambda_K$:
\begin{equation}
\mathbf{C} \boldsymbol{\nu}_i = \lambda_i \boldsymbol{\nu}_i .
\end{equation}
Each principal component $\boldsymbol{\nu}_i$ represents a spatial pattern of charge radius deviations across the nuclear chart.
The eigenvalue $\lambda_i$ quantifies its statistical significance, with $\boldsymbol{\nu}_1$ capturing the dominant variation mode.
These modes reveal common features, model-specific biases, and physics-driven correlations, etc.

The overlap of the principal component $\boldsymbol{\nu}_i$ with the original charge radius model $R_j$ is given by
\begin{equation}
  \frac{\mathbf{R}_j \cdot \boldsymbol{\nu}_i}{\sqrt{|\mathbf{R}_j||\boldsymbol{\nu}_i|}} =
  \frac{\sum_{k=1}^m R_j^k \nu_i^k}{\sqrt{\sum_{k=1}^m (R_j^k)^2} \sqrt{\sum_{k=1}^m (\nu_i^k)^2}}.
\end{equation}

These extracted PCs of nuclear charge radius models can be recombined
to construct new charge radius models. The superposition coefficients are determined
by the overlaps of these PCs with the experimental charge radius data.
Note that when the Hilbert space is reduced to account for the experimental data,
the PCs $\boldsymbol{\nu}_i$ are no longer orthogonal to each other.
Therefore, they are re-orthogonalized using Schmidt orthogonalization
before calculating their overlaps with the experimental data.

\subsection{Power-moderated mean method}

In the PMM method, the consistency of the data is checked by calculating the
following $\chi^2$ value first
\begin{equation}\label{eq:chi}
\chi^2 = \frac{1}{K - 1}\sum_{i = 1}^{K}\frac{(R_i - R_{\mathrm{mp}})^2}{u_i^2 + s^2},
\end{equation}
where $K$ denotes the number of models and $R_{\mathrm{mp}}$ represents the Mandel-Paule (MP) mean, i.e.,
\begin{equation}\label{eq:rmp}
R_{\mathrm{mp}} = \left. \sum_{i = 1}^{K}\frac{R_i}{u_i^2 + s^2} \middle/ \sum_{i = 1}^{K}\frac{1}{u_i^2 + s^2} \right.,
\end{equation}
using $s^2 = 0$ as the initial value. The data are considered consistent if $\chi^2 \leqslant 1$.
When $\chi^2 > 1$, the additional variance $s^2$ in Eqs.~(\ref{eq:chi}) and (\ref{eq:rmp})
must be increased to account for excess uncertainty.
The $\chi^2$ calculation is then iterated via Eq.~(\ref{eq:chi}) until $\chi^2 = 1$ is achieved.

The next step computes the characteristic uncertainty per datum
using the variance corresponding to the larger of
the arithmetic mean $\overline{R}$ or the MP mean $R_{\mathrm{mp}}$, i.e.,
\begin{equation}\label{eq:s}
S = \sqrt{K\cdot\max \bigl( u^2(\overline{R}),\, u^2(R_{\mathrm{mp}}) \bigr)},
\end{equation}
in which
\begin{equation}\label{eq:u2r}
u^2 (\overline{R}) = \sum_{i = 1}^{K}\frac{(R_i - \overline{R})^2}{K(K - 1)},\qquad
\overline{R} = \frac{1}{K}\sum_{i = 1}^{K}R_i,
\end{equation}
and
\begin{equation}\label{eq:u2}
u^2 (R_{\mathrm{mp}}) = \left(\sum_{i = 1}^{K}\frac{1}{u_i^2 + s^2}\right)^{-1}.
\end{equation}
The reference value $R_{\mathrm{ref}}$ and its associated uncertainty $u(R_{\mathrm{ref}})$
are then calculated via a power-moderated weighted mean, using
\begin{eqnarray}
R_{\mathrm{ref}} &=& \sum_{i = 1}^{K} w_{i} R_{i}, \label{eq:rref}\\
\frac{1}{u^2(R_{\mathrm{ref}})} &=& \sum_{i = 1}^{K}
\left[ (\sqrt{u_i^2 + s^2})^{\alpha} \, S^{2 - \alpha} \right]^{-1}, \label{eq:uref}
\end{eqnarray}
where the normalized weighting factor is defined as
\begin{equation}\label{eq:wi}
w_{i} = u^{2}(R_{\mathrm{ref}}) \left[ (\sqrt{u_{i}^{2} + s^{2}})^{\alpha} \, S^{2 - \alpha} \right]^{-1},
\end{equation}
The exponent $\alpha$ in the weighting factor should be selected based on the reliability of the uncertainties $u_i$.
When the uncertainties $u_i$ are uninformative, one should employ the arithmetic mean ($\alpha = 0$).
When the uncertainties $u_i$ are accurate, the weighted mean should be used ($\alpha = 2$).
However, practical uncertainties are typically partially informative yet systematically underestimated;
in such cases, $\alpha$ should be moderated via the heuristic formula $\alpha=2-3/K$
as recommended in Refs.~\cite{Pomme2012, Pomme2015_Metrologia52-S200},
where $K$ denotes the number of models.

Extreme data should be identified using statistical tools prior to determining
the final reference value and uncertainty via the PMM method.
An extreme datum is identified when
\begin{equation} \label{eq:ei}
|e_i| > k \, u(e_i),
\end{equation}
where $e_i=R_i-R_{\mathrm{ref}}$ denotes the difference between an individual value $R_i$
and the candidate reference value $R_{\mathrm{ref}}$.
The coverage factor $k$ typically takes a default value of $k = 2.5$ as recommended
in Refs.~\cite{Pomme2012, Pomme2015_Metrologia52-S200}.
The variance $u^2(e_i)$ is derived from modified uncertainties using normalized weighting factors via
\begin{eqnarray}
u^2 (e_i) &=& u^2 (R_{\mathrm{ref}})\left(\dfrac{1}{w_i} \pm 1\right), \label{eq:uei1}
\end{eqnarray}
where the plus or minus operator in the above formula is selected based on
whether $R_i$ is excluded in or included from the mean, respectively.
When extreme data are identified by the statistical criteria in Eqs.~(\ref{eq:ei}) and (\ref{eq:uei1}),
the data consistency should be re-evaluated using Eqs.~(\ref{eq:chi}) and (\ref{eq:rmp}),
followed by recalculation of the reference value $R_{\mathrm{ref}}$ and its uncertainty $u(R_{\mathrm{ref}})$
via Eqs.~(\ref{eq:s})-(\ref{eq:wi}).
Subsequent extreme values are then detected by reapplying Eqs.~(\ref{eq:ei}) and (\ref{eq:uei1}).
This iterative process continues until no further extreme data require exclusion.
The final reference value $R_{\mathrm{ref}}$ from Eq.~(\ref{eq:rref}) and its uncertainty $u(R_{\mathrm{ref}})$
from Eq.~(\ref{eq:uref}) constitute the PMM method's charge radius prediction and associated standard uncertainty.

\section{Numerical details}\label{sec:details}

The following nuclear models are adopted in these model averaging methods:
(i) The isospin dependent $A^{1/3}$ formula
$R_c=r_A \left[1-b(N-Z)/A\right]A^{1/3}$~\cite{Nerlo-Pomorska1993_ZPA344-359}
with the parameter $r_A$=1.282~fm and $b=0.342$ fitted by experimental data (further denoted by $A^{1/3}$); (b) the relativistic continuum Hartree-Bogoliubov (RCHB) theory~\cite{Xia2018_ADNDT121-122-1};
(iii) the Hartree-Fock-Bogoliubov (HFB) mass model HFB25~\cite{Goriely2013_PRC88-024308};
(iv) the Weizs\"acker-Skyrme (WS) model WS$^\ast$~\cite{Li2021_ADNDT140-101440}; and
(v) the HFB25$^\ast$ model~\cite{Li2021_ADNDT140-101440}.
Note that, by taking into account the nuclear shell corrections ($\Delta E$),
quadrupole and hexadecapole deformations ($\beta_2$ and $\beta_4$) obtained from the WS
and HFB25 models, the following nuclear rms charge radius formula was proposed in Ref.~\cite{Li2021_ADNDT140-101440},
\begin{eqnarray}
r_{c} &=& \sqrt{\frac{3}{5}} \left[ r_{0}A^{1/3} + r_{1}A^{-2/3} + r_{s}I(1-I) + r_{d}\frac{\Delta E}{A} \right] \nonumber\\
&& \times \left[ 1 + \frac{5r_{\beta}}{8\pi}(\beta_{2}^{2} + \beta_{4}^{2}) \right] \label{eq:li2021}
\end{eqnarray}
where $A$ is the mass number and $I$ is the isospin asymmetry.
The five coefficients $r_0$, $r_1$, $r_s$, $r_d$, and $r_\beta$ are determined by fitting to the experimental data,
and their values for the two models are those given in Ref.~\cite{Li2021_ADNDT140-101440}.
In the present work, the fourth and fifth models are denoted as WS$^\ast$ and HFB25$^\ast$, respectively.
The 1013 experimental charge radii with $Z \ge 8$ used in the model averaging methods
are taken from Refs.~\cite{Angeli2013_ADNDT99-69, Li2021_ADNDT140-101440}.
There are 1014 data with $Z \ge 8$ in total, but $^{17}$Ne is unbound in the HFB25 model,
so its charge radius is not considered in these model averaging approaches.

When using PCA to extract PCs, the overlap of the five nuclear charge radius models includes
8022 nuclei; therefore, each vector $\mathbf{R}$, as well as each PC, is a vector in 8022-dimension Hilbert space.
Because five nuclear charge radius models are considered,
there are only 5 irrelevant components.
Therefore, one can obtain 5 PCs extracted from these charge radius models.
The 1013 experimental charge radii are adopted to evaluate the weight of each PC.

\section{Results and discussion}\label{sec:result}

\begin{figure*}[!]
\centering
\includegraphics[width=0.8\textwidth]{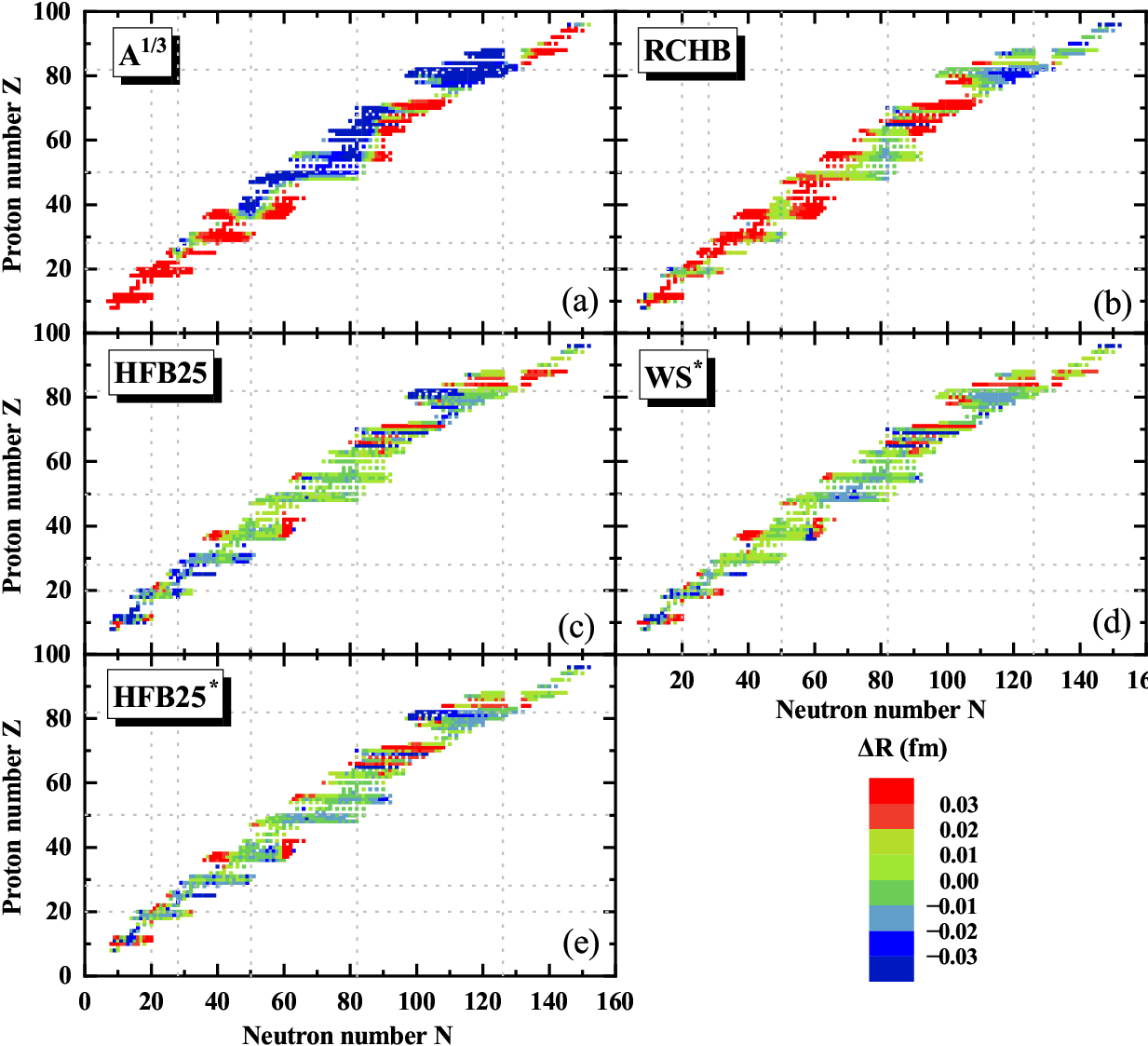}
\caption{Charge radius differences $\Delta R$ between the experimental data~\cite{Angeli2013_ADNDT99-69, Li2021_ADNDT140-101440}
and theoretical predictions for (a) $A^{1/3}$ formula, (b) RCHB, (c) HFB25, (d) WS$^\ast$, and (e) HFB25$^\ast$.
The magic numbers are indicated by grey dashed lines.
}
\label{fig:dev}
\end{figure*}

Figure~\ref{fig:dev} presents the charge radius differences $\Delta R$ between the
experimental data~\cite{Angeli2013_ADNDT99-69, Li2021_ADNDT140-101440} and the predictions of the
five charge radius models under consideration.
The magic numbers are indicated by grey dashed lines.
Large deviations are observed for light nuclei with proton and neutron numbers below 20
in all of these five models, which can be attributed to the cluster structures prevalent in this mass region,
an effect that these models generally fail to capture.
Among the five models, the $A^{1/3}$ formula yields small $\Delta R$ values only in very restricted regions.
The regions of overestimation and underestimation are clearly separated,
suggesting that isospin effects are not adequately accounted for in this simple formula.
The RCHB theory performs well mainly around magic numbers, where nuclei are spherical or weakly deformed.
Large deviations are observed between two adjacent magic numbers.
To date, the deformed relativistic Hartree-Bogoliubov theory in the continuum has only been applied to
even-even nuclei~\cite{Zhang2020_PRC102-024314, Zhang2022_ADNDT144-101488}.
Extending such calculations to all nuclei across the nuclear chart would likely
further improve the description of charge radii.
The other three models (HFB25, WS$^\ast$, and HFB25$^\ast$) exhibit broadly similar patterns,
yet they show notable differences in certain nuclear regions.
For instance, on the proton-rich side of $Z=82$, both HFB25 [Fig.~\ref{fig:dev}(c)] and
HFB25$^\ast$ [Fig.~\ref{fig:dev}(e)] produce large deviations,
whereas WS$^\ast$ [Fig.~\ref{fig:dev}(d)] agrees remarkably well with the data.
It also can be seen that for isotopes with $Z=65$, 69, and 71, all three models (HFB25, WS$^\ast$, and HFB25$^\ast$)
show large deviations for most isotopes.
However, the RCHB theory gives quite small deviations at $Z=69$ [Fig.~\ref{fig:dev}(b)],
and the $A^{1/3}$ formula does so on the neutron-rich side [Fig.~\ref{fig:dev}(a)].
Another interesting feature is that for Cm ($Z=96$) isotopes, the other four models all deviate substantially
from the experimental values, while the $A^{1/3}$ formula reproduces the data rather well.
These observations suggest that each model has its own local strengths and weaknesses.
Therefore, if model averaging methods are designed to account for these local patterns in the deviations
of individual charge radius models, they could effectively reduce the overall discrepancies and
lead to the construction of more precise new models.
In the following, we will compare the performance of the five model averaging methods introduced in Sec.~\ref{sec:theor}.

\begin{figure}[!]
\centering
\includegraphics[width=1.0\columnwidth]{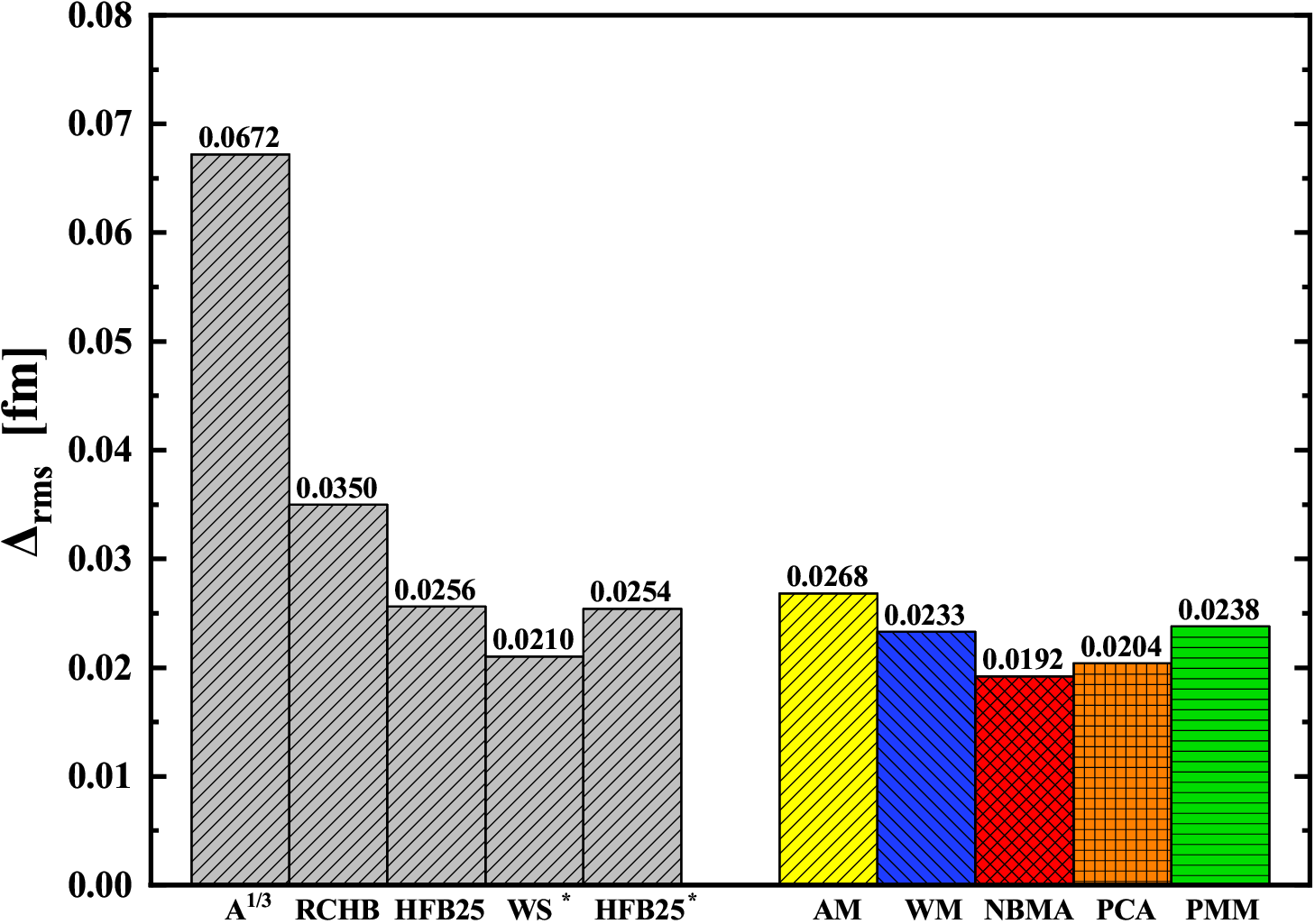}
\caption{The rms deviations ($\Delta_{\rm rms}$) between the experimental data and the predictions of different
charge radius models, together with the results obtained by different model averaging methods.
}
\label{fig:rms}
\end{figure}

First, we will compare the overall performance of these five model averaging methods in nuclear charge radius predictions.
The rms deviations between experimental data and the predictions of five different models,
together with the results obtained by the five model averaging methods are displayed in Fig.~\ref{fig:rms}.
Among the five nuclear charge radius models, WS$^\ast$ exhibits the lowest rms deviation.
Only the NBMA and PCA methods yield a lower rms deviation than WS$^\ast$,
and among the averaging methods, NBMA provides the lowest rms deviations overall.
This is because NBMA assigns nucleus-dependent weights to different models,
with the weights determined by the charge radius accuracy of each model for each
isotopic and isotonic chain.
Therefore, the NBMA method can take into account the advantages of different models in the
isotopic and isotonic chains, and improve the nuclear charge radius prediction by the model averaging method.
In contrast, the AM method gives the largest rms deviation among the averaging methods,
while the WM and PMM methods yield similar rms deviations.
This similarity arises because, in the experimentally known region where the theoretical predictions of
different nuclear models are consistent within uncertainties,
the PMM prediction approaches the WM prediction~\cite{Zhang2024_PRC110-044307}.

\begin{figure*}[!]
\centering
\includegraphics[width=0.8\textwidth]{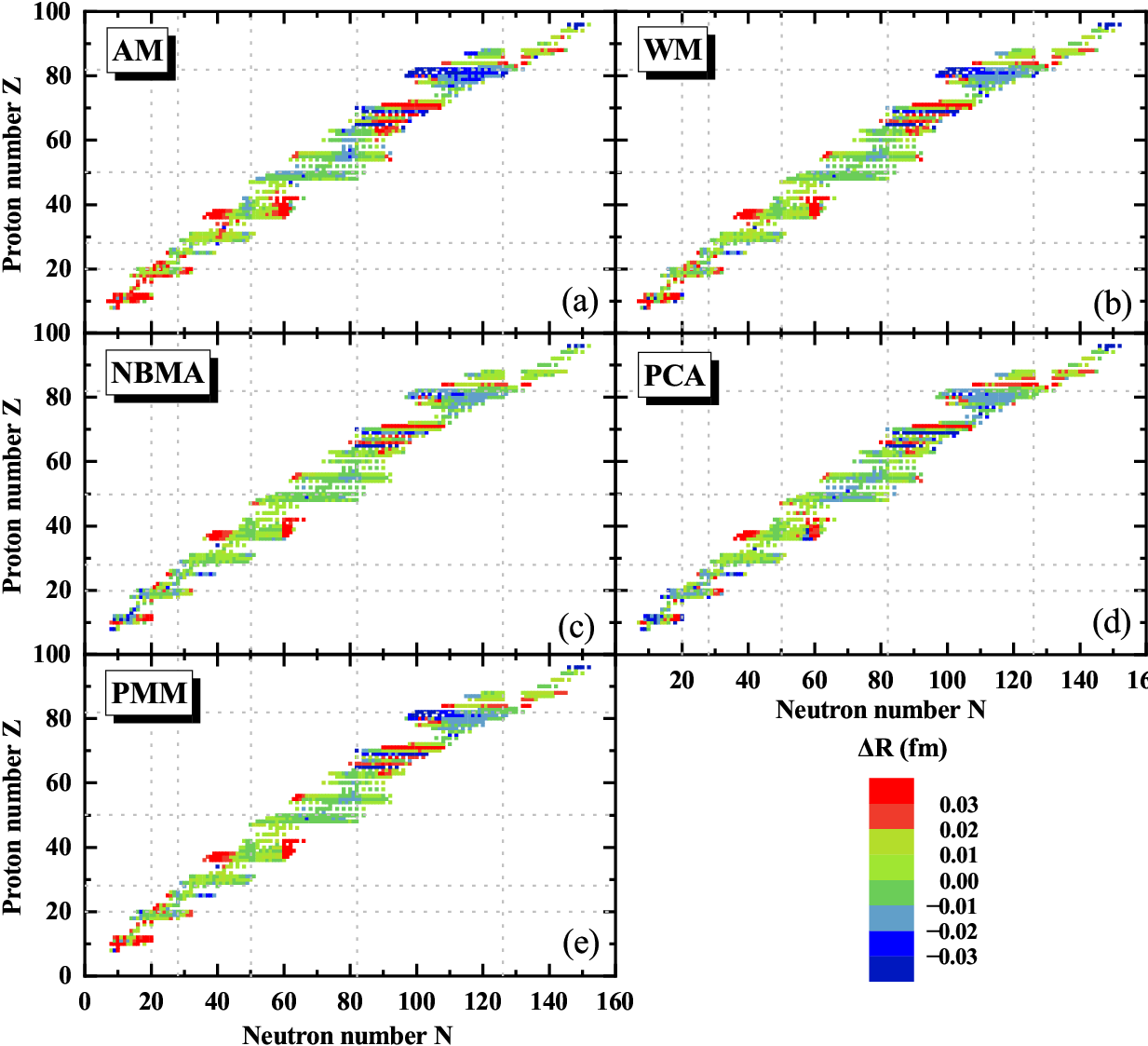}
\caption{Charge radius differences $\Delta R$ between the experimental data~\cite{Angeli2013_ADNDT99-69, Li2021_ADNDT140-101440}
and the predictions obtained with the model averaging methods for (a) AM, (b) WM, (c) NBMA, (d) PCA, and (e) PMM methods.
The magic numbers are indicated by grey dashed lines.
}
\label{fig:mav}
\end{figure*}

\begin{figure}[!]
\centering
\includegraphics[width=1.0\columnwidth]{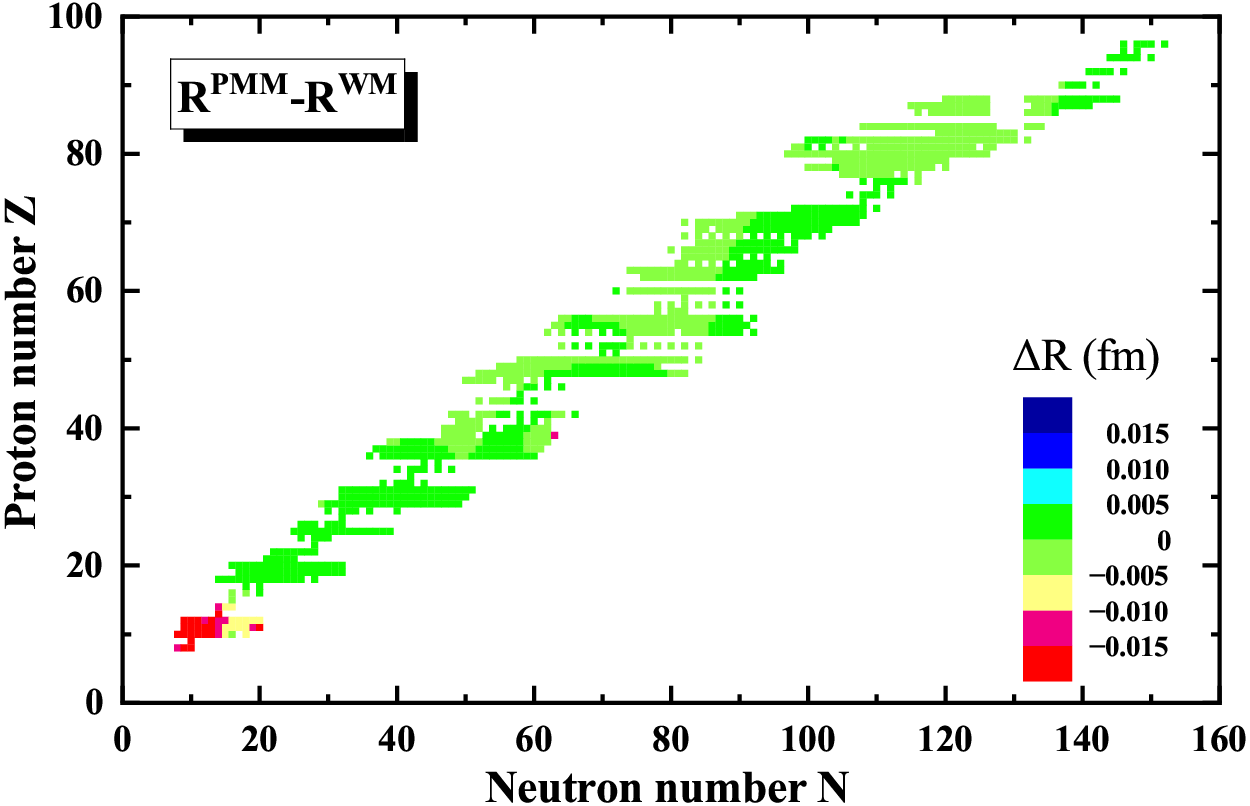}
\caption{The differences between the results obtained by the
PMM method $R^{\rm PMM}$ and WM method $R^{\rm WM}$.
}
\label{fig:pmm-wm}
\end{figure}

Figure~\ref{fig:mav} presents the charge radius differences $\Delta R$ between
the experimental data~\cite{Angeli2013_ADNDT99-69, Li2021_ADNDT140-101440}
and the predictions of the five model averaging methods.
As shown in Fig.~\ref{fig:mav}(a), the AM method yields quite large deviations.
In contrast, the WM method assigns different weights to different models according to their prediction accuracies,
thereby improving the nuclear radius predictions considerably.
As seen in Fig.~\ref{fig:mav}(b), the most significant improvements occur for light nuclei and those around $Z=80$.
As mentioned above, the NBMA method combines the advantages of different models across regions by
assigning a larger weight $P(Z, N|M^k)$ to the model that better reproduces the charge radii
for the isotopic and isotonic chains with the same $Z$ and $N$ as the target nucleus.
Consequently, the NBMA method further enhances the accuracy of charge radius predictions,
as shown in Fig.~\ref{fig:mav}(c).
In particular, the charge radii for isotopes with $Z=78$, 79, and 80 are substantially improved.
The PCA results are quite similar to those of NBMA, except in some specific regions, such as isotopes with $Z=69$ and 71.
The PMM method exhibits a pattern quite similar to that of the WM method.
To quantify this, Fig.~\ref{fig:pmm-wm} displays the differences between the results of the PMM method ($R^{\rm PMM}$)
and the WM method ($R^{\rm WM}$).
It is evident that, except for the lightest nuclei region, nearly all absolute deviations between
these two methods are less than 0.005~fm.
This confirms that the PMM predictions approach those of the WM in regions where experimental data are available.

\begin{figure*}[!]
\centering
\includegraphics[width=0.8\textwidth]{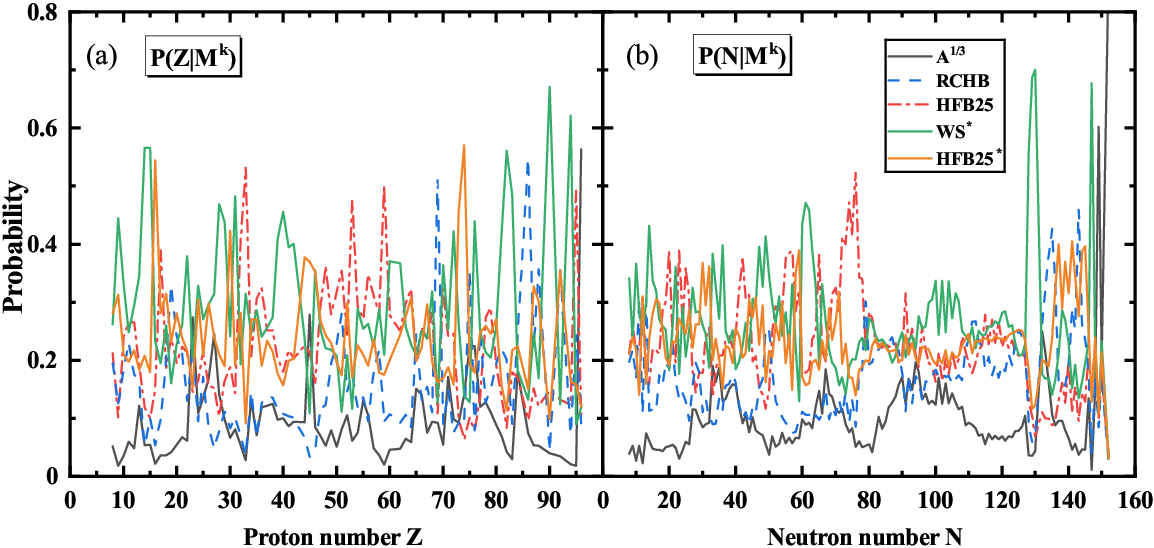}
\caption{Probabilities $P(Z|M^k)$ and $P(N|M^k)$ of the five nuclear charge radius models.
}
\label{fig:pzn1}
\end{figure*}

\begin{figure*}[!]
\centering
\includegraphics[width=0.8\textwidth]{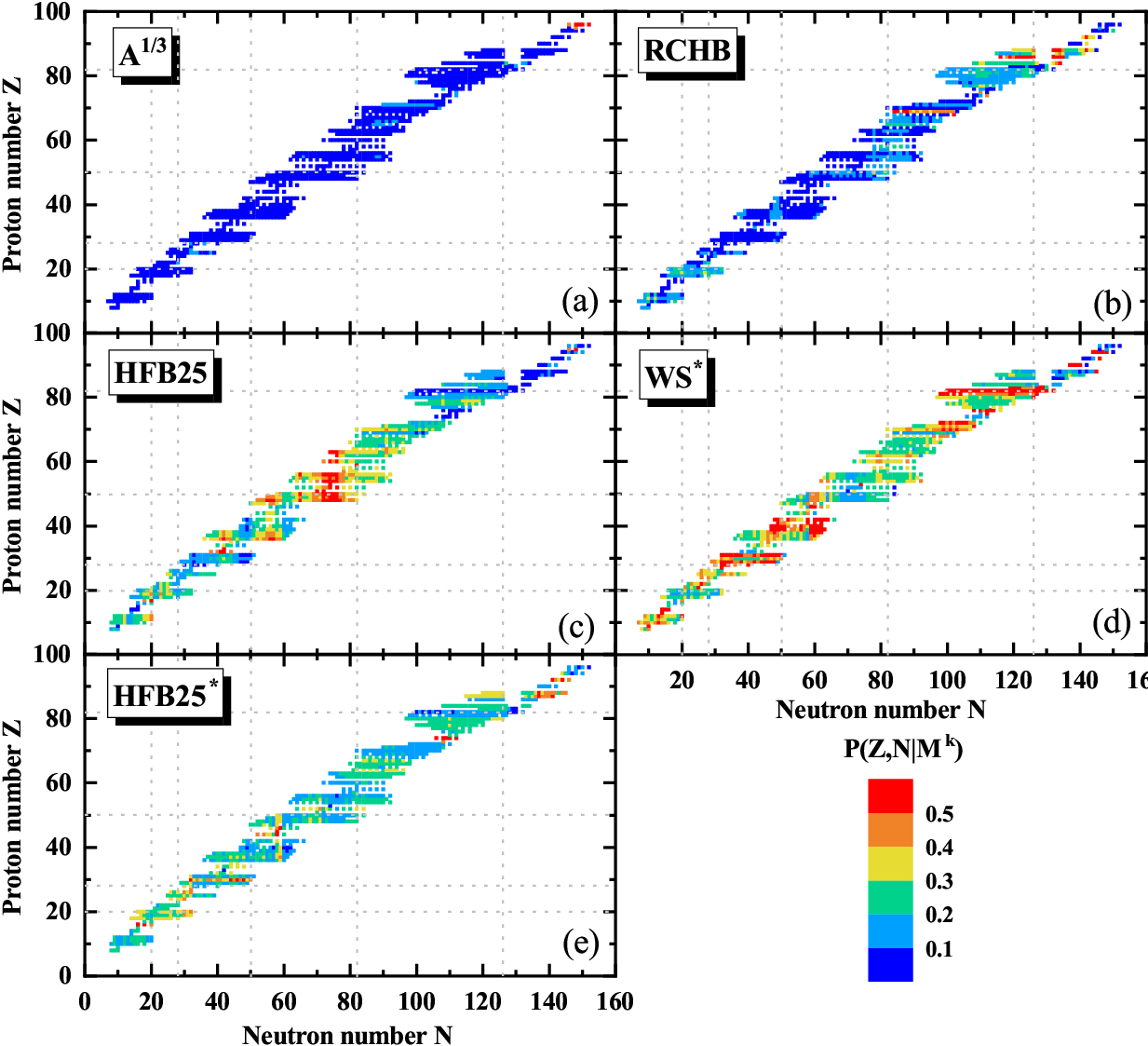}
\caption{Probabilities $P(Z, N|M^k)$ for (a) $A^{1/3}$ formula, (b) RCHB, (c) HFB25, (d) WS$^\ast$, and (e) HFB25$^\ast$.
The magic numbers are indicated by grey dashed lines.
}
\label{fig:pzn2}
\end{figure*}

The probabilities $P(Z|M^k)$ and $P(N|M^k)$ are the key factors in determining
the nuclear charge radius predictions in the NBMA method.
Figure~\ref{fig:pzn1} shows the corresponding probabilities $P(Z|M^k)$ and $P(N|M^k)$
for the five nuclear charge radius models.
It can be seen obviously that the $A^{1/3}$ formula has the lowest probabilities
for most isotopic and isotonic chains since it has the largest rms deviation.
However, it has the largest probabilities for $Z=96$ isotopic and $N=149$, 152 isotonic chains.
This is because all the other four calculations
have very large deviations from the experimental data in Cm isotopes.
Since deformation effects are not included in the spherical RCHB calculations,
the probabilities are large only at the region close to the shell and sub-shell closures.
The dominant probabilities only appear at $Z=69, 86$ isotopic and $N=135, 143$ isotonic chains.
The other three models have their own advantages at certain isotopic and isotonic chains.
Therefore, the NBMA method can improve the nuclear charge radius prediction by
taking into account the advantages of different models in the isotopic and isotonic chains.

To provide a clearer view of the probability distribution, Fig.~\ref{fig:pzn2} displays the
probabilities $P(Z, N|M^k)$ assigned to each of the five nuclear charge radius models for every nucleus.
Since the $A^{1/3}$ formula shows large deviations over nearly the entire nuclear chart,
its probabilities $P(Z, N|M^k)$ are dominant only for the Cm isotopes.
In contrast, the RCHB model attains relatively large probabilities only around the magic number region,
with dominant probabilities appearing for isotopes with $Z=69$, 86, and 88.
The other three models (HFB25, WS$^\ast$, and HFB25$^\ast$) all have relatively large probabilities
across most of the nuclear chart, especially WS$^\ast$.
It is worth noting that although HFB25 and HFB25$^\ast$ have quite similar rms deviations,
their probability patterns $P(Z, N|M^k)$ differ considerably.
The dominant probabilities of HFB25 are sparse and scattered in only a few regions,
whereas those of HFB25$^\ast$ are predominant in most isotopes of the medium-mass region.
This difference may arise from the fact that HFB25$^\ast$ and WS$^\ast$ share the same
phenomenological formula for charge radii, but employ different deformations, shell energies, and coefficients.
Moreover, the rms deviation of WS$^\ast$ is much smaller than that of HFB25$^\ast$,
which likely renders HFB25$^\ast$ less favored than WS$^\ast$ across the entire nuclear chart.
Therefore, the NBMA method can partially account for the local structures in model deviations,
thereby effectively reducing the local discrepancies between experimental charge radii and model predictions.
By assigning different weights to different nuclear models in different regions,
the NBMA method is able to combine the strengths of these models and is expected to
achieve better predictive capability for nuclear charge radii.

\begin{table*}[!]
\centering
\caption{\label{tab:pca}
Eigenvalues of the five PCs extracted from the five charge radius models and the overlaps of the five PCs
with the five charge radius models and the experimental data CR21 from Refs.~\cite{Angeli2013_ADNDT99-69, Li2021_ADNDT140-101440}.}

\begin{tabular*}{1.0\textwidth}{@{\extracolsep{\fill}}lccccc}
\toprule[1pt]\specialrule{0em}{1pt}{1pt}
Models & PC1 & PC2 & PC3 & PC4 & PC5 \\
\hline
Eigenvalues    & $1.51\times10^{2}$ & $7.89\times10^{-3}$ & $1.57\times10^{-3}$ & $5.53\times10^{-4}$ & $2.04\times10^{-4}$ \\
\hline
$A^{1/3}$      & 0.999894    &  0.014522   &  0.000556  &  0.000585  &  0.000119  \\
RCHB           & 0.999970    & -0.004866   &  0.006033  &  0.000177  &  0.000008  \\
HFB25          & 0.999992    & -0.001060   & -0.001607  & -0.003426  & -0.000837  \\
WS$^{\ast}$    & 0.999984    & -0.004106   & -0.002586  &  0.002475  & -0.001338  \\
HFB25$^{\ast}$ & 0.999987    & -0.003901   & -0.002396  &  0.000191  &  0.002047  \\
CR21           & 0.999985    & -0.002970   & -0.000887  &  0.001053  & -0.001337  \\
\bottomrule[1pt]
\end{tabular*}
\end{table*}

The PCA approach builds a new model by recombining the principal components to achieve higher precision.
Table~\ref{tab:pca} shows the eigenvalues of the five PCs extracted from the five charge radius models,
together with the overlaps between each PC and the individual models as well as
the experimental data CR21 from Refs.~\cite{Angeli2013_ADNDT99-69, Li2021_ADNDT140-101440}.
The eigenvalues drop by nearly six orders of magnitude from PC1 ($\lambda_1 = 1.51\times10^{2}$)
to PC5 ($\lambda_5 = 2.04\times10^{-4}$),
indicating that the first component overwhelmingly dominates the variance among these five models.
PC1 shows overlaps exceeding 0.9999 with all models and with CR21,
confirming that it captures the universal $A^{1/3}$ trend,
which is the leading bulk contribution to nuclear sizes.
The higher PCs reveal finer physical distinctions.
In PC2, the $A^{1/3}$ formula has a positive overlap, while all the other models have negative overlaps.
This sign separation indicates that PC2 mainly encodes the general microscopic structural effects,
such as shell and pairing correlations, which are absent in the simple phenomenological formula
but present in all macroscopic-microscopic and mean-field calculations.
A more striking pattern emerges in PC3, where the overlaps split strictly according to
the presence or absence of deformation: the $A^{1/3}$ formula and spherical RCHB model
have positive overlaps, whereas all three deformed models exhibit negative overlaps.
Because the overall sign of a PC is arbitrary in PCA, the physical significance lies in the relative sign across models.
Here, the clear sign reversal identifies PC3 as the component representing deformation effects.
Note that the overlap of the experimental data CR21 with PC3 is negative,
closely matching the sign of the deformed models and differing from the spherical ones.
This consistency demonstrates that the PCs extracted purely from theoretical models are physically meaningful.
The remaining PCs, PC4 and PC5, have even smaller eigenvalues and show less systematic overlap patterns,
suggesting that they correspond to residual, model specific details rather than universal physical effects.
The PCA decomposition thus provides a natural hierarchy of charge radius predictions: volume (PC1),
general microscopic structure, such as shell efffects (PC2), and deformation (PC3).
This framework not only helps to interpret the model differences but also offers
a systematic way to construct improved empirical models by recombining these orthogonal components,
analogous to the successful application to nuclear mass prediction in Ref.~\cite{Wu2024_SSPMA67-272011}.

\begin{figure}[h]
\centering
\includegraphics[width=1.0\columnwidth]{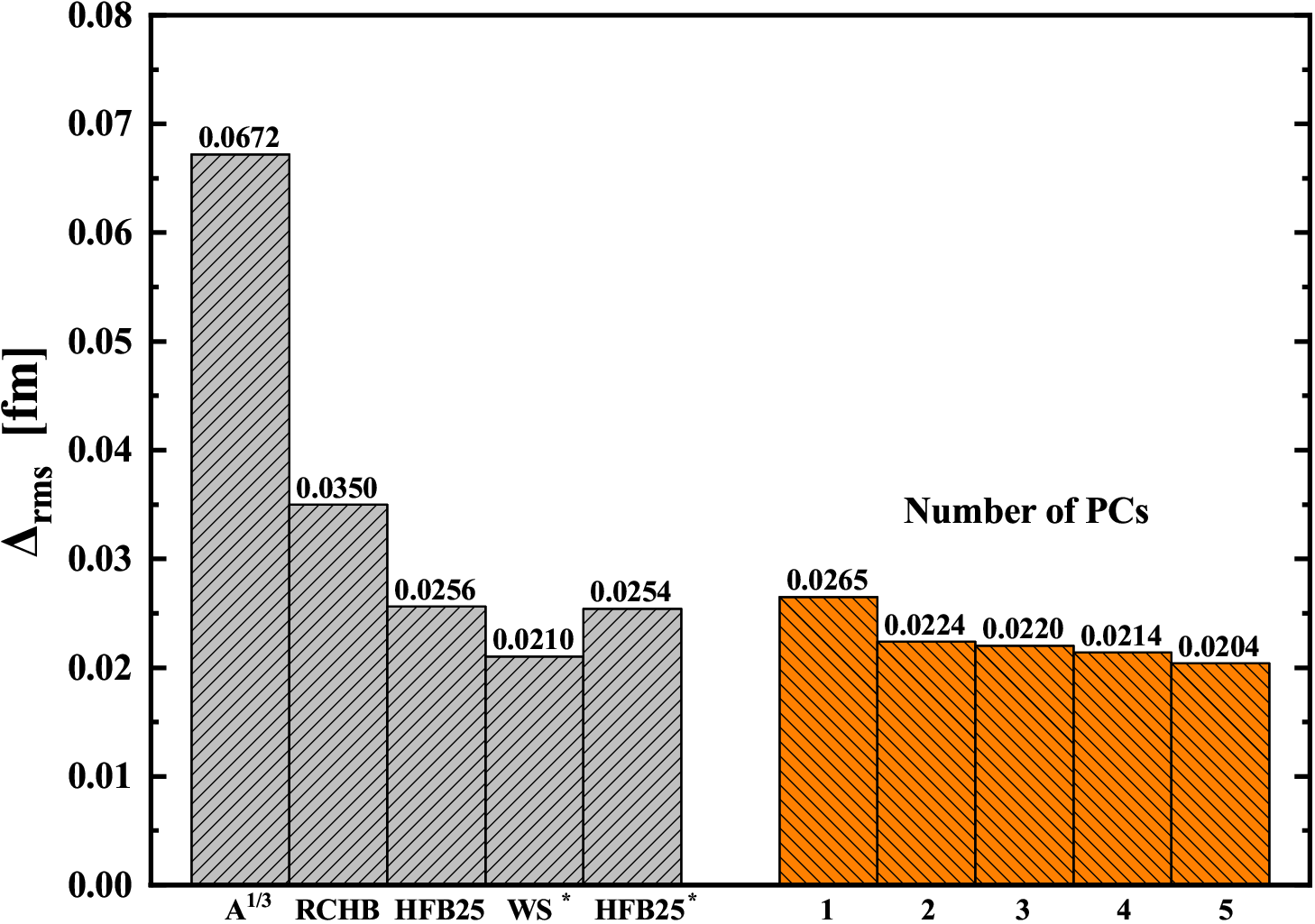}
\caption{The rms deviations ($\Delta_{\rm rms}$) between the experimental data and the predictions of different
charge radius models, together with the results obtained by PCA with different numbers of PCs.
}
\label{fig:pca}
\end{figure}

Figure~\ref{fig:pca} shows the rms deviations ($\Delta_{\rm rms}$) between the experimental data
and the predictions of various charge radius models,
as well as those for the new models reconstructed from different numbers of PCs.
As the number of included PCs increases from 1 to 5,
the rms deviation monotonically decreases from 0.0265~fm to 0.0204~fm.
This trend indicates that each additional PC brings in valuable physical information that
improves the agreement with experimental data.
When only PC1 is included, the rms deviation already reaches a precision of 0.0265~fm.
The largest reduction occurs when moving from PC1 to PC2 (a decrease of 0.0041~fm),
implying that PC2 captures the most important correction beyond the global volume trend.
With only these two PCs included, the results are already very close to those obtained by WS$^{\ast}$,
and significantly better than all the other original models.
The improvements contributed by PC3, PC4, and PC5 are progressively smaller,
suggesting that these components represent increasingly subtle or model-specific corrections.
Nevertheless, the continuous decline in the rms deviation demonstrates that the PCA decomposition
successfully extracts orthogonal patterns from the theoretical models,
and that recombining these patterns can systematically enhance predictive power.
Importantly, the final rms deviation of 0.0204~fm achieved with all five PCs is notably
smaller than that of any single original model,
confirming that the collaborative use of multiple models via PCA yields superior accuracy.

\begin{figure}[h]
\centering
\includegraphics[width=1.0\columnwidth]{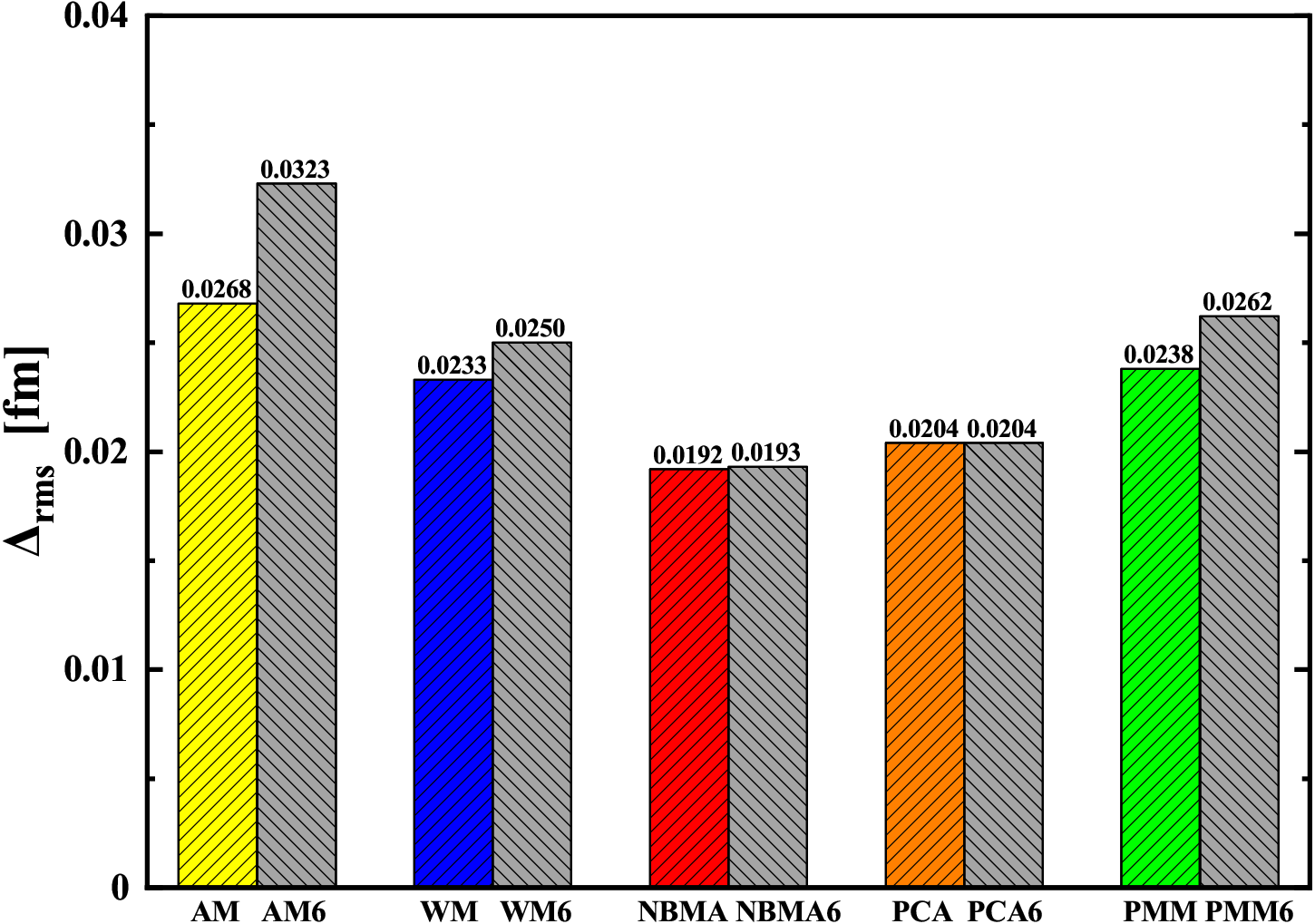}
\caption{The rms deviations ($\Delta_{\rm rms}$) between the experimental data and the predictions of different
model averaging methods. The name of these model averaging methods followed by ``6'' means the results obtained
by the $Z^{1/3}$ law (6 nuclear charge radius models in total) are considered in these approaches.
}
\label{fig:av6}
\end{figure}

To test the robustness of these model averaging methods when additional datasets are included,
the $Z^{1/3}$ formula~\cite{Zeng1957_APS13-357}
\begin{equation} \label{eq:rz}
R_c=r_Z Z^{1/3}
\end{equation}
is added to the previous five models, where $Z$ is the proton number
and $R_c=\sqrt{\frac{5}{3}}\langle r^2 \rangle^{1/2}$ with $\langle r^2 \rangle^{1/2}$ denoting the rms charge radius.
The parameter $r_Z$ is determined to be 1.640~fm by fitting the 1013 experimental data.
The rms deviation of the $Z^{1/3}$ formula is 0.0744~fm, which is larger than those of the other five models.
After adding the results of this model, the model averaging methods are applied to the 6 datasets,
denoted as AM6, WM6, etc., in Fig.~\ref{fig:av6}, and the results are compared with the original ones.
It can be seen that all results become worse with the $Z^{1/3}$ formula except for the PCA method.
The PCA6 results are the same as the original PCA with five principal components, even when a model with a quite large rms deviation is considered.
If 5 decimal places are chosen, the results actually improve.
This is because PCA can extract the new effects from the $Z^{1/3}$ formula and make them contribute to
improving nuclear charge radius predictions.
Therefore, developing new theoretical charge radius models that include novel physical effects
remains highly beneficial, even when their direct agreement with experimental data is poor,
since PCA can isolate these effects and convert them into improvements in the predictive power for nuclear charge radii.
It also can be seen that the rms deviation of NBMA6 is only slightly larger than that of NBMA.
This is because the weights in NBMA are determined on the basis of Bayes' theorem,
and the weights for a model with larger rms deviations are quite small.

\begin{figure}[!]
\centering
\includegraphics[width=1.0\columnwidth]{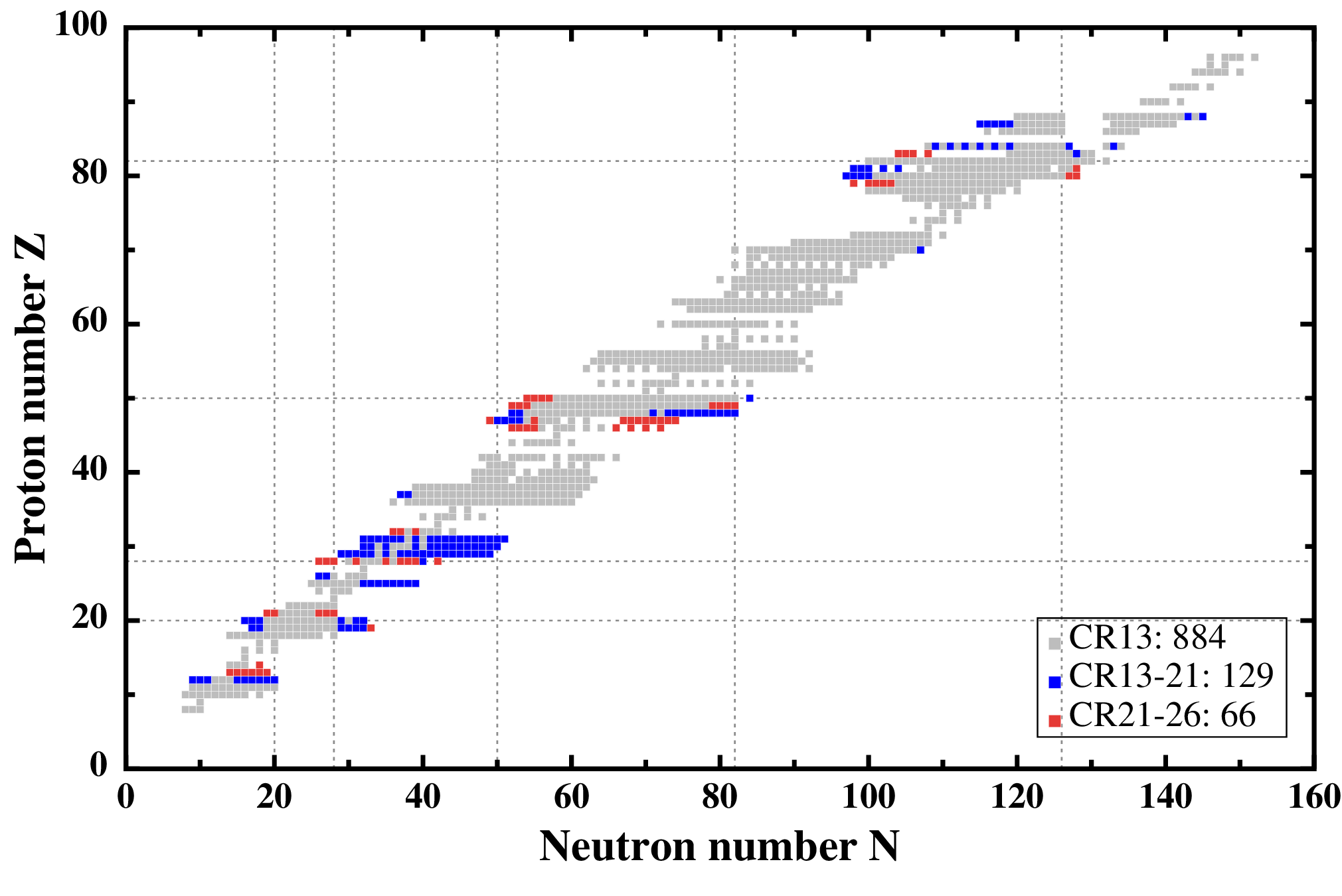}
\caption{Nuclei in the training set (CR13) and test sets (CR13-21 and CR21-26)
for examining the extrapolation power of the five nuclear models and the model averaging methods.
}
\label{fig:nc}
\end{figure}

The predictive performance of these nuclear charge radius models and model averaging methods
is evaluated using the rms deviations for both the training (CR13) and test (CR13-21) sets.
The 1013 experimental charge radii in Ref.~\cite{Li2021_ADNDT140-101440} with $Z \ge 8$
are divided into two subsets: the training set CR13 of 884 nuclei appearing in Ref.~\cite{Angeli2013_ADNDT99-69},
and the test set CR13-21 (test1) of 129 nuclei first reported in Ref.~\cite{Li2021_ADNDT140-101440}.
In addition, 66 newly measured charge radii CR21-26 (test2) after year 2021,
which are not included in Refs.~\cite{Angeli2013_ADNDT99-69, Li2021_ADNDT140-101440},
are also used to test the predictive performance of the models.
The separation of the training and test sets can be seen in Fig.~\ref{fig:nc} in detail.
The 66 new data are shown in Table~\ref{tab:nd},
including the mean square charge radius difference $\delta\langle r^2\rangle^{\rm exp}$,
the rms charge radius $r_{\rm c}^{\rm exp}$, and the corresponding references.
Note that these 66 new data are not included in the fitting process for the $A^{1/3}$ formula,
WS$^{\ast}$, and HFB25$^{\ast}$ models.
Therefore, they represent genuine extrapolations for these three models.

\begin{longtable}{@{\extracolsep{\fill}} cccccc}
\caption{\label{tab:nd} The 66 newly observed data CR21-26 (test2) after year 2021, which
are not compiled in Ref.~\cite{Angeli2013_ADNDT99-69, Li2021_ADNDT140-101440},
including the mean square charge radius difference $\delta\langle r^2\rangle^{\rm exp}$,
the rms charge radius $r_{\rm c}^{\rm exp}$, and the corresponding references.}\\
\toprule[1pt]\specialrule{0em}{1pt}{1pt}
$Z$ & $N$ & $A$ & $\delta\langle r^2\rangle^{\rm exp}$ (fm$^2$) & $r_{\rm c}^{\rm exp}$ (fm) & Ref. \\
\hline
\endfirsthead
\caption*{\tablename\ \thetable\ (Continued)}\\
\toprule[1pt]
\specialrule{0em}{1pt}{1pt}
$Z$ & $N$ & $A$ & $\delta\langle r^2\rangle^{\rm exp}$ (fm$^2$) & $r_{\rm c}^{\rm exp}$ (fm) & Ref. \\
\hline
\endhead
\bottomrule[1pt]
\endfoot
13 & 15 & 28 & 0.003  & 3.0615 & \cite{Heylen2021_PRC103-014318} \\
   & 16 & 29 & 0.141  & 3.0839 & \cite{Heylen2021_PRC103-014318} \\
   & 17 & 30 & 0.164  & 3.0877 & \cite{Heylen2021_PRC103-014318} \\
   & 18 & 31 & 0.301  & 3.1098 & \cite{Heylen2021_PRC103-014318} \\
   & 19 & 32 & 0.120  & 3.0805 & \cite{Heylen2021_PRC103-014318} \\
\hline
14 & 18 & 32 & 0.195  & 3.1535 & \cite{Koenig2024_PRL132-162502} \\
\hline
19 & 33 & 52 & 0.790  & 3.5481 & \cite{Koszorus2021_NatPhys17-439} \\
\hline
21 & 19 & 40 & -0.226 & 3.5139 & \cite{Koenig2023_PRL131-102501} \\
   & 20 & 41 & -0.305 & 3.5026 & \cite{Koenig2023_PRL131-102501} \\
   & 26 & 47 & -0.137 & 3.5265 & \cite{Bai2025_PRL134-182501} \\
   & 27 & 48 & -0.239 & 3.5120 & \cite{Bai2025_PRL134-182501} \\
   & 28 & 49 & -0.286 & 3.5053 & \cite{Bai2025_PRL134-182501} \\
\hline
28 & 26 & 54 & -0.522 & 3.7366 & \cite{Sommer2022_PRL129-132501} \\
   & 27 & 55 & -0.607 & 3.7252 & \cite{Sommer2022_PRL129-132501} \\
   & 28 & 56 & -0.626 & 3.7226 & \cite{Sommer2022_PRL129-132501} \\
   & 31 & 59 & -0.180 & 3.7823 & \cite{Malbrunot-Ettenauer2022_PRL128-022502} \\
   & 35 & 63 & 0.277  & 3.8422 & \cite{Malbrunot-Ettenauer2022_PRL128-022502} \\
   & 37 & 65 & 0.385  & 3.8562 & \cite{Malbrunot-Ettenauer2022_PRL128-022502} \\
   & 38 & 66 & 0.493  & 3.8702 & \cite{Malbrunot-Ettenauer2022_PRL128-022502} \\
   & 39 & 67 & 0.514  & 3.8729 & \cite{Malbrunot-Ettenauer2022_PRL128-022502} \\
   & 42 & 70 & 0.806  & 3.9105 & \cite{Malbrunot-Ettenauer2022_PRL128-022502} \\
\hline
32 & 36 & 68 & -0.184 & 4.0363 & \cite{Wang2024_PLB856-138867} \\
   & 37 & 69 & -0.244 & 4.0288 & \cite{Wang2024_PLB856-138867} \\
   & 39 & 71 & -0.223 & 4.0314 & \cite{Wang2024_PLB856-138867} \\
\hline
46 & 52 & 98 & -1.231 & 4.4192 & \cite{Geldhof2022_PRL128-152501} \\
   & 53 & 99 & -1.121 & 4.4316 & \cite{Geldhof2022_PRL128-152501} \\
   & 54 & 100 & -0.929 & 4.4532 & \cite{Geldhof2022_PRL128-152501} \\
   & 55 & 101 & -0.827 & 4.4646 & \cite{Geldhof2022_PRL128-152501} \\
   & 66 & 112 & 0.361  & 4.5957 & \cite{Geldhof2022_PRL128-152501} \\
   & 68 & 114 & 0.487  & 4.6094 & \cite{Geldhof2022_PRL128-152501} \\
   & 70 & 116 & 0.574  & 4.6189 & \cite{Geldhof2022_PRL128-152501} \\
   & 72 & 118 & 0.648  & 4.6268 & \cite{Geldhof2022_PRL128-152501} \\
\hline
47 & 49 & 96  & -1.21  & 4.4292 & \cite{Reponen2021_NC12-4596} \\
   & 55 & 102 & -0.67  & 4.4898 & \cite{Reponen2021_NC12-4596} \\
   & 67 & 114 & 0.384  & 4.6057 & \cite{Reponen2021_NC12-4596} \\
   & 68 & 115 & 0.454  & 4.6133 & \cite{Reponen2021_NC12-4596} \\
   & 69 & 116 & 0.500  & 4.6183 & \cite{Reponen2021_NC12-4596} \\
   & 70 & 117 & 0.568  & 4.6256 & \cite{Reponen2021_NC12-4596} \\
   & 71 & 118 & 0.607  & 4.6298 & \cite{Reponen2021_NC12-4596} \\
   & 72 & 119 & 0.675  & 4.6372 & \cite{Reponen2021_NC12-4596} \\
   & 73 & 120 & 0.715  & 4.6415 & \cite{Reponen2021_NC12-4596} \\
   & 74 & 121 & 0.767  & 4.6471 & \cite{Reponen2021_NC12-4596} \\
\hline
49 & 52 & 101 & -1.274 & 4.4755 & \cite{Karthein2024_NatPhys20-1719} \\
   & 53 & 102 & -1.171 & 4.4870 & \cite{Vernon2025_PRC111-064325} \\
   & 54 & 103 & -1.019 & 4.5039 & \cite{Karthein2024_NatPhys20-1719} \\
   & 79 & 128 & 0.545  & 4.6743 & \cite{Vernon2025_PRC111-064325} \\
   & 81 & 130 & 0.642  & 4.6846 & \cite{Vernon2025_PRC111-064325} \\
   & 80 & 129 & 0.598  & 4.6799 & \cite{Karthein2024_NatPhys20-1719} \\
   & 82 & 131 & 0.645  & 4.6850 & \cite{Karthein2024_NatPhys20-1719} \\
\hline
50 & 54 & 104 & -1.497 & 4.5121 & \cite{Gustafsson2025_PRL135-222501} \\
   & 55 & 105 & -1.408 & 4.5219 & \cite{Gustafsson2025_PRL135-222501} \\
   & 56 & 106 & -1.257 & 4.5386 & \cite{Gustafsson2025_PRL135-222501} \\
   & 57 & 107 & -1.180 & 4.5470 & \cite{Gustafsson2025_PRL135-222501} \\
   & 83 & 133 & 0.406  & 4.7182 & \cite{Gustafsson2025_PRL135-222501} \\
\hline
79 & 98  & 177 & -0.990 & 5.3452 & \cite{Cubiss2023_PRL131-202501} \\
   & 100 & 179 & -0.796 & 5.3634 & \cite{Cubiss2023_PRL131-202501} \\
   & 101 & 180 & -0.274 & 5.4118 & \cite{Cubiss2023_PRL131-202501} \\
   & 102 & 181 & -0.203 & 5.4184 & \cite{Cubiss2023_PRL131-202501} \\
   & 103 & 182 & -0.184 & 5.4199 & \cite{Cubiss2023_PRL131-202501} \\
\hline
80 & 127 & 207 & 0.503 & 5.4923 & \cite{Goodacre2021_PRL126-032502} \\
   & 128 & 208 & 0.624 & 5.5033 & \cite{Goodacre2021_PRL126-032502} \\
\hline
81 & 128 & 209 & 0.330 & 5.5060 & \cite{Yue2024_PRC110-034315} \\
\hline
83 & 104 & 187 & -0.949 & 5.4345 & \cite{Barzakh2021_PRL127-192501} \\
   & 105 & 188 & -0.335 & 5.4907 & \cite{Barzakh2021_PRL127-192501} \\
   & 106 & 189 & -0.859 & 5.4428 & \cite{Barzakh2021_PRL127-192501} \\
   & 108 & 191 & -0.810 & 5.4473 & \cite{Barzakh2021_PRL127-192501} \\
\bottomrule[1pt]
\end{longtable}

\begin{table}[h]
\caption{\label{t:1}
The rms deviations of the training set CR13 ($\Delta_{\rm rms}^{\rm train}$) and test sets
CR13-21 ($\Delta_{\rm rms}^{\rm test1}$) and CR21-26 ($\Delta_{\rm rms}^{\rm test2}$)
for different nuclear charge radius models and model averaging methods.}
\begin{tabular*}{1.0\columnwidth}{@{\extracolsep{\fill}}cccc}
			\toprule[1pt]\specialrule{0em}{1pt}{1pt}
			Model          & $\Delta_{\rm rms}^{\rm train}$~(fm) & $\Delta_{\rm rms}^{\rm test1}$~(fm) & $\Delta_{\rm rms}^{\rm test2}$~(fm) \\
			\hline
			$A^{1/3}$      &  0.0652   &  0.0722   & 0.0618    \\
			RCHB           &  0.0361   &  0.0260   & 0.0281   \\
			HFB25          &  0.0254   &  0.0260   & 0.0266   \\
			WS$^{\ast}$    &  0.0212   &  0.0177   & 0.0203   \\
			HFB25$^{\ast}$ &  0.0257   &  0.0232   & 0.0233   \\
			\hline
			AM             &  0.0271   &  0.0227   &  0.0170   \\
			WM             &  0.0237   &  0.0187   &  0.0165  \\
			NBMA           &  0.0196   &  0.0170   &  0.0183  \\
			PCA            &  0.0207   &  0.0167   &  0.0195  \\
			PMM            &  0.0242   &  0.0199   &  0.0158  \\
			\bottomrule[1pt]
		\end{tabular*}
\end{table}

Table~\ref{t:1} lists the rms deviations for the training set ($\Delta_{\rm rms}^{\rm train}$)
and two test sets ($\Delta_{\rm rms}^{\rm test1}$ and $\Delta_{\rm rms}^{\rm test2}$) for each model and averaging method.
Among the individual models, the simple $A^{1/3}$ law clearly performs worst on both
test sets ($\Delta_{\rm rms}^{\rm test1}=0.0722$~fm, $\Delta_{\rm rms}^{\rm test2}=0.0618$~fm),
confirming that a global mass-number and isospin dependence alone cannot capture local structural variations.
The microscopic or phenomenological models, namely RCHB, HFB25, WS$^{\ast}$, and HFB25$^{\ast}$,
all yield substantially lower rms deviation on both test sets.
HFB25 gives rms deviation of $0.0260$ and $0.0266$~fm for test1 and test2, respectively,
while its phenomenological variant HFB25$^{\ast}$ improves these values to $0.0232$ and $0.0233$~fm,
indicating that the phenomenological adjustment enhances extrapolation performance,
albeit with a slight increase in training error ($0.0254$ to $0.0257$~fm).
WS$^{\ast}$ stands out as the best individual model, with the rms deviation of
$0.0177$ and $0.0203$~fm for test1 and test2, respectively,
even outperforming its refined counterpart HFB25$^{\ast}$.
This suggests that the particular parametrization of WS$^{\ast}$ offers superior extrapolation capability.

Turning to the model averaging methods, all of them yield competitive results,
generally surpassing the performance of most individual models on both test sets.
Among them, PCA achieves the lowest rms deviation for test1 ($0.0167$~fm), closely followed by NBMA ($0.0170$~fm)
and PMM ($0.0199$~fm). The excellent performance of PCA on test1 indicates that principal component analysis
can effectively extract the dominant correlated features among individual models, yielding an optimal linear combination.
The NBMA method, which relies on Bayesian weighting, also performs very well, confirming that assigning smaller
weights to less reliable models helps reduce systematic biases.
However, on the independent test2 set, PMM provides the smallest rms deviation ($0.0158$~fm),
while NBMA and PCA give slightly larger values ($0.0183$ and $0.0195$~fm, respectively),
even exceeding WM ($0.0165$~fm) and AM ($0.0170$~fm).
This observation suggests that PMM, whose design automatically adjusts data uncertainties
to achieve a smooth transition from WM to AM in regions with scarce experimental information,
exhibits superior robustness when facing truly unknown data.
In contrast, NBMA and PCA, while excellent on the test1 set that shares similar coverage
with the training data, may be more sensitive to the distribution shift present in test2.
The WM method, which relies on inverse-variance weighting, performs quite
well on both test sets ($0.0187$ and $0.0165$~fm) and even outperforms some more
sophisticated approaches on test2, indicating that simple variance-based weighting remains a reliable baseline.
The AM, albeit the simplest, gives rms deviations of $0.0227$ and $0.0170$~fm, for test1 and test2, respectively,
showing that a naive equal-weight combination already provides a substantial improvement over the worst individual model.
Another noteworthy feature is the behaviour of training versus test errors.
For AM, WM, and PMM, the rms deviations decrease monotonically from train to test1 to test2,
which is a reassuring sign of generalisation. In contrast, NBMA and PCA show rms deviations
that are slightly larger than their test1 values on test2, although they remain well within competitive ranges.
This may reflect that the weights optimised on the training set,
while effective for the data distribution of test1, are not fully transferable to the newer test2 data.
Nevertheless, the absolute deviations of all averaging methods on test2 are below $0.020$~fm,
which is remarkable given the wide diversity of the underlying individual models.
These results demonstrate that model averaging offers a balanced and reliable strategy for nuclear charge radius predictions,
combining high accuracy on familiar data with robust extrapolation to new measurements.
The choice of the optimal averaging method may depend on the specific application scenario:
PCA or NBMA is preferred when the test set is expected to resemble the training distribution,
whereas PMM is more suitable for genuine predictions in unexplored regions.

\begin{figure*}[!]
\centering
\includegraphics[width=0.8\textwidth]{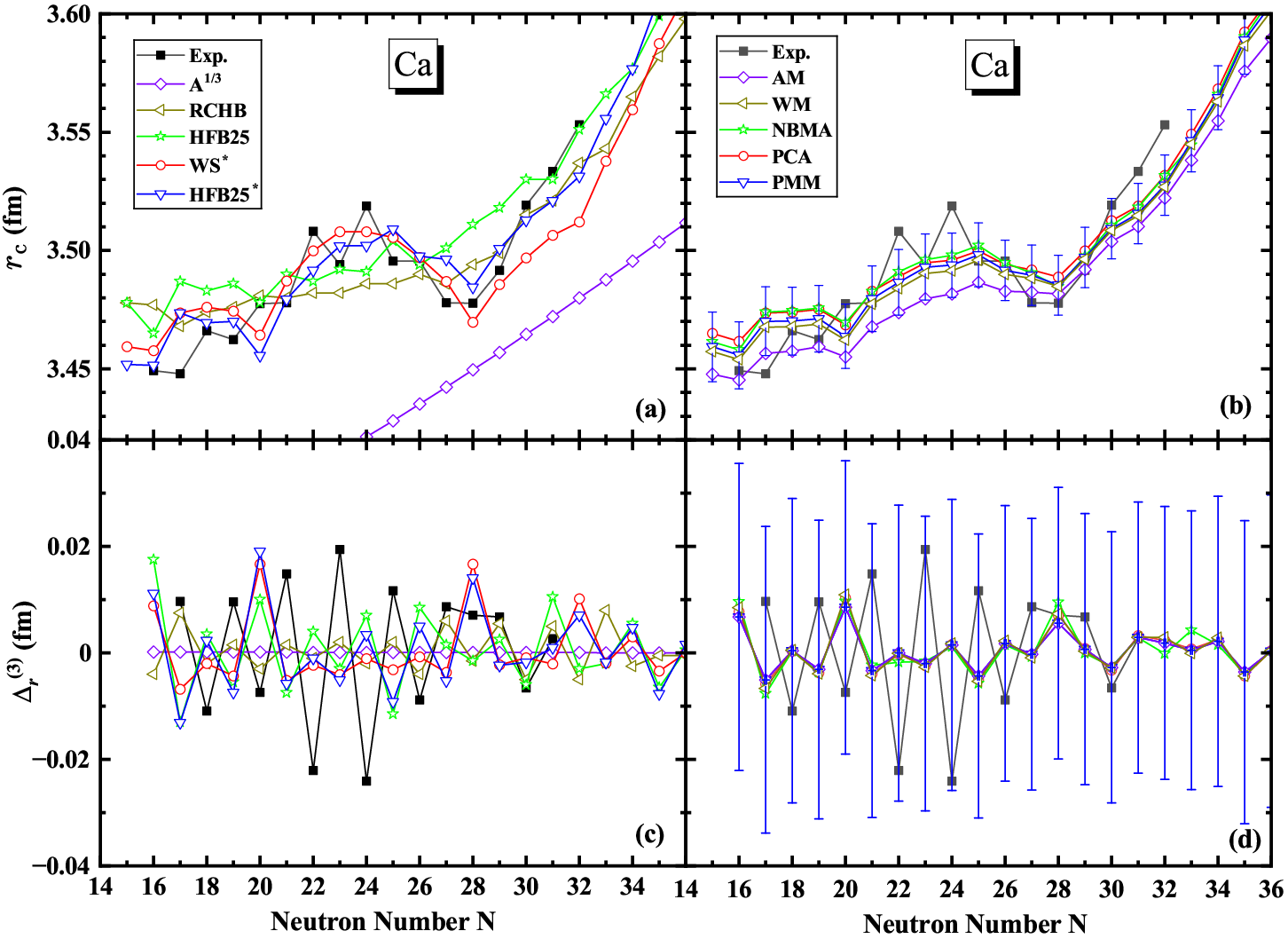}
\caption{Rms charge radii and odd-even staggering of Ca isotopes obtained by
different nuclear charge radius models and model averaging methods.
}
\label{fig:ca}
\end{figure*}

The evolution of charge radii of calcium isotopes is of particular interest~\cite{Ruiz2016_NatPhys12-594,
Miller2019_NatPhys15-432}, as it can provide a stringent test of nuclear charge radius models.
The charge radii of $^{40}$Ca and $^{48}$Ca are quite close to each other, resulting in an inverse parabola
in this region, which is followed by an unexpectedly large increase in the $N>28$ region.
Furthermore, strong odd-even staggering (OES) is observed in the $N>28$ region,
reflecting the fine structure of the charge radii in calcium isotopes.
The OES parameter for the charge radii is defined as
\begin{equation}\label{eq:d3}
\Delta_r^{(3)}(Z,N)=\frac{1}{2}[r(Z,N-1) - 2r(Z,N) + r(Z,N+1)] \ ,
\end{equation}
where $r(Z, N)$ is the rms charge radius for a nucleus with proton number $Z$ and neutron number $N$.

To examine the predictive power of these model averaging methods,
we compare the rms charge radii and OES of Ca isotopes
predicted by different nuclear charge radius models and model averaging methods, as shown in Fig.~\ref{fig:ca}.
It can be seen in Fig.~\ref{fig:ca}(a) that only WS$^{\ast}$ and HFB25$^{\ast}$ models can reproduce
the inverse parabola structure between $^{40}$Ca and $^{48}$Ca,
as well as the large increase of charge radii in the $N>28$ region.
Although HFB25$^{\ast}$ can reproduce the inverse parabola structure, the HFB25 model cannot.
This suggests that the phenomenological formula given in Ref.~\cite{Li2021_ADNDT140-101440}
provides valuable physical effects.
The simple $A^{1/3}$ law gives the worst results.
Turning to the OES predictions, Fig.~\ref{fig:ca}(c) shows that only the RCHB theory
can reproduce the trend of the experimental OES.
However, the amplitude of the calculated OES is much less pronounced compared with the experimental data.
The $A^{1/3}$ formula exhibits no OES throughout the entire isotopic chain,
and the WS$^{\ast}$ model shows quite weak OES except at the $N=20$ and 28 shell closures.
The OES in the HFB25 and HFB25$^{\ast}$ models are somewhat stronger,
but they are still weak compared with the data.
Note that although OES can be obtained in the WS$^{\ast}$, HFB25, and HFB25$^{\ast}$ models,
the phases of the calculated OES are opposite to the experimental data.

As for the results of the model averaging methods, Fig.~\ref{fig:ca}(b) shows that
all the averaged results are quite similar, except for the AM method,
which gives obviously smaller rms charge radii for all isotopes.
Nevertheless, its trend remains consistent with the other averaging methods.
The parabola structure is reproduced by all these methods, albeit with a slight underestimation.
This indicates that it is difficult to extract the fine structure using model averaging alone.
It can also be seen that, after considering uncertainties, most of the experimental data can be reproduced by the PMM method.
Turning to the OES predictions, Fig.~\ref{fig:ca}(d) shows that all the model averaging methods can produce OES,
but the phases of the calculated OES are still opposite to the experimental data.
Although the RCHB theory can give the correct phase of the OES, the amplitude of the calculated OES is too small.
The model averaging methods fail to capture this effect.
This indicates that if nearly all underlying models fail to reproduce the fine structure of nuclear charge radii,
the model averaging methods cannot correct this deficiency either.
Therefore, it is very important to incorporate physical insights when constructing nuclear models.

\section{Summary}\label{sec:summary}

In summary, we have systematically investigated the performance of five model averaging methods,
namely, the arithmetic mean (AM), weighted mean (WM), naive Bayesian model averaging (NBMA),
principal component analysis (PCA), and power-moderated mean (PMM) methods in predicting nuclear charge radii
based on five commonly used nuclear models: the isospin-dependent $A^{1/3}$ formula,
relativistic continuum Hartree-Bogoliubov theory, Hartree-Fock-Bogoliubov (HFB) model HFB25,
the Weizs\"acker-Skyrme (WS) model WS$^\ast$, and HFB25$^{\ast}$.
The results demonstrate that the NBMA method yields the smallest root-mean-square (rms) deviation
among all methods, effectively reducing local discrepancies by exploiting regional model advantages.
The PCA method gives a slightly larger rms deviation than NBMA, but remains highly competitive.
It not only provides accurate predictions but also offers valuable physical insights
by recombining principal components, enabling a systematic construction of improved empirical models.
The rms deviation obtained by the PCA method remains
nearly unchanged even when adding the $Z^{1/3}$ formula, which has a larger rms deviation than other models,
whereas the results of other methods become worse.
The PMM method shows robust performance, smoothly interpolating between the WM and AM behaviors
in known and unknown regions, respectively, while providing reasonable uncertainty estimates
and automatically adjusting data uncertainties to achieve consistency.

Furthermore, by dividing the 1013 experimental data into training and test sets,
the extrapolation capabilities of these methods are validated.
Moreover, 66 newly observed data after year 2021 are also used to test the extrapolation capabilities.
It is found that model averaging offers a reliable strategy for nuclear charge radius predictions,
combining high accuracy on familiar data with robust extrapolation to new measurements.
The choice of the optimal averaging method may depend on the specific application scenario:
PCA and NBMA are preferred when the test set is expected to resemble the training distribution,
whereas PMM is more suitable for predictions in unexplored regions.
In addition, the rms charge radii and odd-even staggering patterns of calcium isotopes are briefly discussed,
further illustrating the applicability of the model averaging framework.
The results indicate that if nearly all underlying models fail to reproduce the fine structure of nuclear charge radii,
the model averaging methods cannot correct this deficiency either.
This study highlights the potential of model averaging techniques,
particularly NBMA, PMM, and PCA, to improve nuclear charge radius predictions and provide a
reliable foundation for future theoretical developments.

\begin{acknowledgments}
This work is supported by the National Natural Science Foundation of China (Nos. 11875027, 12475121, 12375109, 12405134),
the Fundamental Research Funds for the Central Universities (2026SL014),
the Anhui project (Z010118169), and the Key Research Foundation of Education Ministry of Anhui Province (2023AH050095),
and the High-Performance Computing Platform of North China Electric Power University.
\end{acknowledgments}


%

\end{document}